# A Theory of Growing Crystalline Nanorods – Mode I

Feng Du and Hanchen Huang *
Department of Mechanical and Industrial Engineering, Northeastern University, Boston, MA 02115, USA

Nanorods grow in two modes through physical vapor deposition (PVD). In mode I, monolayer surface steps dictate the diameter of nanorods. In mode II, multiple-layer surface steps dictate the diameter, which is the smallest possible under physical vapor deposition [X. B. Niu *et al.*, Phys. Rev. Lett **110**, 136102 (2013) and F. Du & H. C. Huang, Phys. Rev. Materials. **1**, 033401 (2017)]. This paper reports closed-form theories of terrace lengths and nanorod diameter during the growth in mode I, as a function of deposition conditions. The accompanying lattice kinetic Monte Carlo simulations verify the theories. This study reveals that (1) quasi-steady growth exists for each set of nanorod growth condition, (2) the characteristic length scales, including terrace lengths and nanorod diameter at the quasi-steady state, depend on the deposition conditions – deposition rate $F$, substrate temperature $T$, and incidence angle $\theta$ – only as a function of $l_{2D}/\tan\theta$, with $l_{2D} = 2\left(\frac{v_{2D}}{F\cos\theta}\right)^{\frac{1}{3}}$ as a diffusion-limited length scale and $v_{2D}$ as the atomic diffusion jump rate over monolayer surface steps.

* Author to whom correspondence should be addressed; e-mail: h.huang@northeastern.edu

# 1. Introduction

The growth of crystalline nanorods through physical vapor deposition (PVD) proceeds in two possible modes. In comparison, both modes rely on the incidence angle being glancing or oblique in experiments [1-4]. In contrast, the growth of mode I relies on monolayer surface steps and that of mode II relies on multiple-layer surface steps to limit the surface diffusion or mass transport [5]; and typically the growth of mode I takes place on a wetting substrate, and that of mode II takes place on a non-wetting substrate [6].

The growth of mode II leads to the smallest diameter of nanorods, because of the large three-dimensional Ehrlich-Schwoebel (3D ES) diffusion barrier over multiple-layer surface steps [7-9]. Driven by the stronger desire of growing smaller nanorods, the theory of nanorod diameter for growth of mode II has been formulated before that of mode I, verified by atomistic simulations, and validated by PVD experiments [5, 10]. This theory, coupled with the theory of nanorod separation [11], has enabled the design of not only small but also well-separated nanorods, and their experimental realization [5]. The availability of small and well-separated metallic nanorods has in turn resulted in the technology of metallic glue [12, 13].

The growth of mode I leads to a larger diameter of nanorods than that of mode II does, because of the smaller two-dimensional Ehrlich-Schwoebel (2D ES) diffusion barrier over monolay-layer surface steps [14, 15]. However, this growth mode bridges with that of thin films, and is therefore scientifically interesting [16, 17]. For thin films, the wedding cake model [18-20] incorporates the effects of 2D ES barriers and builds on the BCF theory [21], but excludes the effects of non-zero incidence angle. Incorporating the effects of non-zero incidence angle, a recent theory shows that the growth of thin film transitions to the nanorod growth of mode I at a critical coverage and above a critical incidence angle [22]. Beyond the transition point, an important characteristic length scale is the diameter of nanorods, and it is the primary focus of this invesitgation.

This paper reports a closed-form theory of the nanorod diameter, in terms of deposition conditions – substrate temperature (or diffusion jump rate), deposition rate, and incidence angle of deposition flux – as well as nanorod separation, which depends on deposition condition and substrate patterns. Further, this paper also reports closed-form theories of terrace lengths, the sum of which defines the nanorod diameter.

# 2. Theory

We first conceptually describe in Section 2.1 the framework of theoretical formulations, in terms of (1) characteristic length scales of interest, (2) quasi-steady state condition, and (3) number of coupled equations vs number of physical unknowns. In Section 2.2, we present theoretical formulations for quasi-steady state growth and numerical results to gain insights of terrace lengths as a function of time. In Section 2.3, we take into account the gained insights to derive approximate and closed-form theories, and numerically show the validity of the approximation. In Section 2.4, we use lattice kinetic Monte Carlo (KMC) simulations to verify the closed-form theories.

The KMC simulations are for the epitaxial growth of a prototype Cu [5, 22, 23]. As a brief recap of the simulations, atoms with one coordination has a diffusion hopping rate of $v_s =$ $v_0 e^{-E_s/kT}$ on flat surfaces, $v_{2D} = v_0 e^{-E_{2D}/kT}$ over monolayer surface steps, and $v_{3D} = v_0 e^{-E_{3D}/kT}$ over multiple-layer surface steps. Here, $v_0$ is $5 \times 10^{11}$ /s, $E_s$ is 0.06 eV, $E_{2D}$ is 0.16 eV and, $E_{3D}$ is 0.40 eV [7, 9] and $kT$ is the Boltzmann factor. As discussed in Ref. [24], in order to have a comparable length scale of surface islands as in three dimensions under typical room temperature of 300 K and typical deposition rate of 1 nm/s, the substrate temperature in two-dimensional simulations needs to be choosen around 100 K.

## 2.1 Conceptual Framework

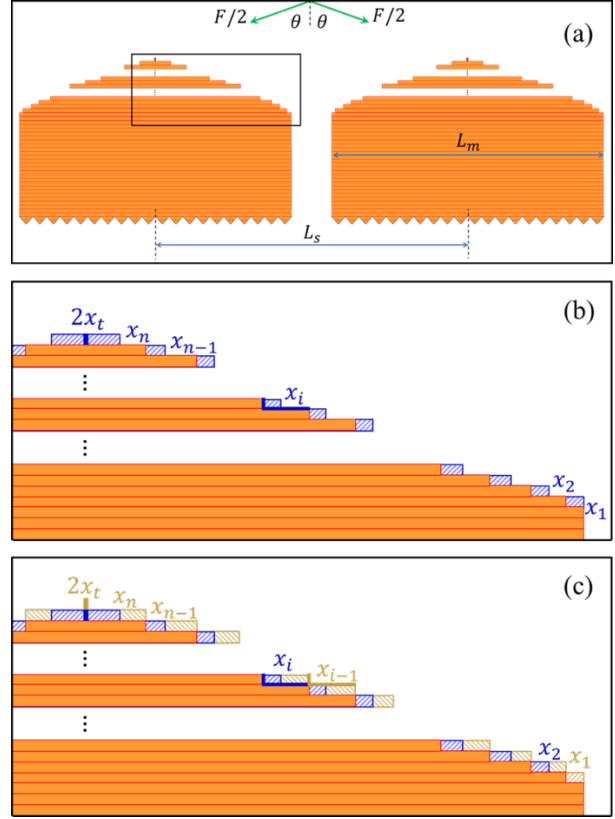

Fig. 1. (a) Schematic of nanorod morphologies at quasi-steady state, (b) surface morphology from time zero shown as solid to $\tau_0$ with newly grown region shown as meshed blue, and (c) surface morphology from time $\tau_0$ to $\tau$ with newly grown region shown as meshed tan. The $i$-th step and the $i$-th terrace of length $x_i$ are marked by solid lines, to illustrate their relative positions.

Based on experimental observations [1-4], and as shown later by KMC simulations, the mode I growth of nanorods will reach a quasi-steady state. Figure 1(a) illustrates the top section of nanorods at quasi-steady state – a wedding cake like top surface is bounded by multiple-layer surface steps on both sides, in two dimensions or 1+1 dimensions. At quasi-steady state, the entire top surface grows taller by one layer during one growth period $\tau$, with the starting and the ending top surface morphologies identical. Shown in Figs. 1(b) and 1(c) are the expanded views of the boxed area of Fig. 1(a). At time zero, a nucleus of mathermatically zero dimension forms on the top; and at time $\tau$, the top surface grows taller by one layer and a nucleus forms again on the top;

Fig. 1(c). During the time period from 0 to $\tau$, the first step advances so that $x_1$ becomes zero at time $\tau_0$; Fig. 1(b). As growth continues with time from $\tau_0$ to $\tau$, the surface morphology returns to that at time zero but the entire surface grows one layer taller; Fig. 1(c).

As shown in Fig. 1(b), during the the time between 0 and $\tau$ there are $n$ monolayer surface steps and $n$ terraces, plus one island above the $n$-th terrace. Accordingly, there are $n+1$ lengths $x_1$, $x_2$, ..., $x_n$, and $x_t$, characterizing the dimensions of the terraces and the top island. There are also $n+1$ rate equations that govern the evolution of $x_1$, $x_2$, ..., $x_n$, and $x_t$ as a function of time. Solution of the $n+1$ rate equations with boundary conditions gives rise to the terrace length $l_1$, $l_2$, ..., and $l_n$ at time zero, in terms of $\tau_0$ and $\tau$; the corresponding $l_t$ is zero. We use the term of boundary condition instead of initial condition here since the condition is not for the start of time. To eliminate $\tau_0$ and $\tau$ in the solutions, two additional equations are necessary. One of the two additional equations is for the critical nucleation size during growth, and the other for the mass conservation. Once the terrace lengths – $l_1$, $l_2$, ..., and $l_n$ – are determined, their sum defines the diameter $L_m$ as $2(l_1 + l_2 ... + l_n)$.

## 2.2 General Theory

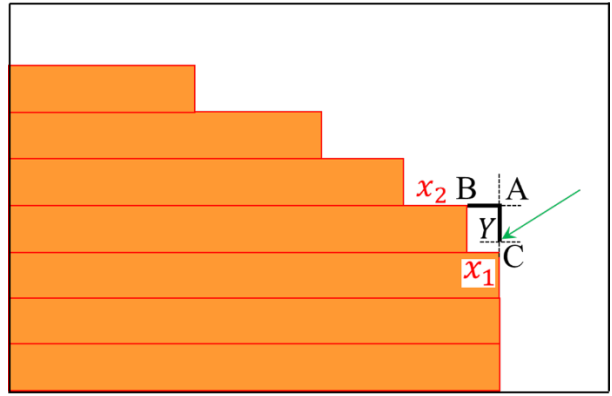

Fig. 2. Expanded view near the first step, showing the lower boundary of trajectories of source atoms (arrowed line) and the shadowing effects during the period of $0 \leq t < \tau_0$.

Our formulations start with the length of the first terrace $x_1$; in this paper, length is in the unit of atomic diameter. At time zero, the lowest trajectory of source atoms cannot reach below the first terrace. Otherwise, the growth would not be in quasi-steady state, because $L_m$ would continue to increase. As shown in Fig. 2, source atoms that go through AB and AC will contribute to the advancement of the first step. The length of AB is simply $x_1$. However, the length of AC is $Y$ and it changes with time $t$. We assume that $Y$ linearly decreases with time and it goes to 0 at $\tau_0$. As we will see later in section 2.3, under quasi-steady state terrace lengths increase linearly with time, and the longer terraces lead to linear decrease of $Y$. That is, $Y = (\tau_0 - t)/\tau$. The initial vaule of $Y$ is now a parameter $\tau_0/\tau$. Ignoring the inter-step diffusion as in the previous formulations [22], we have:

$$\frac{dx_1}{dt} = \frac{1}{2}\left[-F\cos\theta\, x_1 - F\sin\theta\, \frac{(\tau_0-t)}{\tau}\right] \tag{1}$$

Here, we have assumed that the deposition flux from the left does not reach the terraces on the right side of the nanorod. As shown later in Fig. 3, this assumption is valid for all the terraces except the top two. With the boundary condition that $x_1$ is 0 at $t \geq \tau_0$, the solution of Eq. (1) is:

$$\begin{cases} x_1(t) = \frac{2\tan\theta}{\tau F\cos\theta}\left[e^{\frac{F\cos\theta(\tau_0-t)}{2}} - \frac{F\cos\theta(\tau_0-t)}{2} - 1\right] & \text{if } 0 \leq t < \tau_0 \\ x_1(t) = 0 & \text{if } \tau_0 \leq t < \tau \end{cases} \tag{2}$$

At time zero, when a new layer nucleates on the top island, the length of the first terrace is $l_1$:

$$l_1 = \frac{2\tan\theta}{\tau F\cos\theta}\left(e^{\frac{F\cos\theta\tau_0}{2}} - \frac{F\cos\theta\tau_0}{2} - 1\right) \tag{3}$$

It is noted that the length $l_1$ is in terms of deposition parameters ($\theta$ and $F$), as well as two parameters $\tau_0$ and $\tau$. Based on the length of the first terrace, we now determine the length of the second terrace. During one period of growth, the shadowing condition for the second terrace varies between the time $0 \leq t < \tau_0$ and $\tau_0 \leq t < \tau$. During the time interval of $0 \leq t < \tau_0$, the length of the first terrace is none zero and the length of the second terrace $x_2$ increases due to the growth the terrace below (first terrace), and decreases due to deposition on itself (second terrace) $F\cos\theta\, x_2$ and on to the vertical step $F\sin\theta \times 1$. The governing equation of $x_2$ is therefore:

$$\frac{dx_2}{dt} = \frac{1}{2}\left[F\cos\theta\, x_1 + F\sin\theta\, \frac{(\tau_0-t)}{\tau} - (F\cos\theta\, x_2 + F\sin\theta)\right] \tag{4}$$

During the time interval of $\tau_0 \leq t < \tau$, the first terrace no longer exists. So, the terrace length of the second layer decreases due to deposition on itself (second terrace) with length of $x_2$ and on to the vertical step of length $Y'$, which can be derived in the same as for the first terrace; Fig. 2. That is, $Y' = 1 + (\tau_0 - t)/\tau$. The governing equation of $x_2$ during this time period is therefore:

$$\frac{dx_2}{dt} = \frac{1}{2}\left[-F\cos\theta\, x_2 - F\sin\theta\left(1 + \frac{\tau_0}{\tau} - \frac{t}{\tau}\right)\right] \tag{5}$$

With the continuity condition at time $t = \tau_0$ and the boundary condition that $x_2 = l_1$ at $t = \tau$, the solution of $x_2$ is:

$$\begin{cases} x_2(t) = \frac{2\tan\theta}{\tau F\cos\theta}\left[\frac{F\cos\theta(t-\tau_0)}{2}e^{\frac{F\cos\theta(\tau_0-t)}{2}} + e^{\frac{F\cos\theta(\tau+\tau_0-t)}{2}} - \left(\frac{F\cos\theta\tau}{2} + 1\right)\right] & \text{if } 0 \leq t < \tau_0 \\ x_2(t) = \frac{2\tan\theta}{\tau F\cos\theta}\left[e^{\frac{F\cos\theta}{2}(\tau+\tau_0-t)} - \frac{F\cos\theta(\tau+\tau_0-t)}{2} - 1\right] & \text{if } \tau_0 \leq t < \tau \end{cases} \tag{6}$$

At time zero, when a new layer nucleates on the top island, the length of the second terrace is $l_2$:

$$l_2 = \frac{2\tan\theta}{\tau F\cos\theta}\left(e^{\frac{F\cos\theta(\tau+\tau_0)}{2}} - \frac{F\cos\theta\tau_0}{2}e^{\frac{F\cos\theta\tau_0}{2}} - \frac{F\cos\theta\tau}{2} - 1\right) \tag{7}$$

It is noted again that the length $l_2$ is in terms of deposition parameters ($\theta$ and $F$), as well as two parameters $\tau_0$ and $\tau$.

For the 3rd terrace, the governing equation of $x_3$ is:

$$\frac{dx_3}{dt} = \frac{1}{2}F\cos\theta\, (x_2 - x_3) \tag{8}$$

This equation is subject to the boundary condition that $x_3$ becomes $l_2$ at $t = \tau$. The solution for

$x_3$ is:

$$x_3(t) = \frac{\tan\theta}{\tau F \cos\theta} \left[ \left( \begin{array}{c} \frac{F^2 \cos^2\theta}{4} e^{\frac{F\cos\theta\tau_0}{2}} t^2 + F\cos\theta\, e^{\frac{F\cos\theta\tau_0}{2}} \left( e^{\frac{F\cos\theta\tau}{2}} - \frac{F\cos\theta\tau_0}{2} \right) t \\ + \left( \frac{F^2 \cos^2\theta \tau_0^2}{4} - 2 \right) e^{\frac{F\cos\theta\tau_0}{2}} \\ + \left( 2e^{\frac{F\cos\theta(\tau_0+\tau)}{2}} - F\cos\theta(\tau_0+\tau) e^{\frac{F\cos\theta\tau_0}{2}} \right) e^{\frac{F\cos\theta\tau}{2}} \\ +2 + F\cos\theta(\tau_0 - \tau) \\ -(F\cos\theta\,\tau + 2) \end{array} \right) e^{-\frac{F\cos\theta t}{2}} \right] \quad \text{if } 0 \leq t < \tau_0$$

$$x_3(t) = \frac{\tan\theta}{\tau F \cos\theta} \left[ \left( \begin{array}{c} \left( F\cos\theta\, t + 2e^{\frac{F\cos\theta\tau}{2}} - F\cos\theta(\tau_0+\tau) \right) e^{\frac{F\cos\theta\tau_0}{2}} \\ +2 + F\cos\theta(\tau_0 - \tau) \\ -F\cos\theta(\tau_0 + \tau - t) - 4 \end{array} \right) e^{\frac{F\cos\theta(\tau-t)}{2}} \right] \quad \text{if } \tau_0 \leq t < \tau$$

(9)

At time zero, when a new layer nucleates on the top island, the length of the third terrace is $l_3$:

$$l_3 = \frac{\tan\theta}{\tau F \cos\theta} \left[ \begin{array}{c} \left( 2e^{\frac{F\cos\theta(\tau_0+\tau)}{2}} - F\cos\theta(\tau_0+\tau) e^{\frac{F\cos\theta\tau_0}{2}} + F\cos\theta(\tau_0-\tau) + 2 \right) e^{\frac{F\cos\theta\tau}{2}} \\ - \left( 2 - \frac{F^2\cos^2\theta\tau_0^2}{4} \right) e^{\frac{F\cos\theta\tau_0}{2}} - F\cos\theta\,\tau - 2 \end{array} \right] \quad (10)$$

It is noted again that the length $l_3$ is in terms of deposition parameters ($\theta$ and $F$), as well as two parameters $\tau_0$ and $\tau$.

For the $i$-th terrace, with $4 \leq i \leq (n-1)$, the governing equation of $x_i$ is:

$$\frac{dx_i}{dt} = \frac{1}{2} F \cos\theta\, (x_{i-1} - x_i) \tag{11}$$

This equation is subject to the boundary condition that $x_i$ becomes $l_{i-1}$ at $t = \tau$. Although in principle, $x_i$ for $4 \leq i \leq (n-1)$ can be obtained in the same way as $x_1$, $x_2$ and $x_3$, the accurate solution of $x_i$ in closed-form will be more complex than that in Eq. (9) even if it is achievable. Therefore, we will first numerically solve for $x_i$ – together with $x_n$ and $x_t$ – to gain insights of the quasi-steady state growth, and then derive the closed-form solution based on the insights.

The governing equation of $x_n$ is different from that of $x_i$ in Eq. (11), since (1) the direct deposition onto the top surface of $2x_t$ in length contributes to the change of $x_n$, and (2) the direct deposition from left also reaches part of the $n$-th terrace of $x_n - \tan\theta$ in length when $x_n \geq \tan\theta$. As a result, the governing equation of $x_n$ becomes:

$$\begin{cases} \frac{dx_n}{dt} = \frac{1}{2} F\cos\theta\,(x_{n-1} - x_n) - F\cos\theta\, x_t - \frac{1}{2} F\cos\theta\,(x_n - \tan\theta) & \text{if } x_n \geq \tan\theta \\ \frac{dx_n}{dt} = \frac{1}{2} F\cos\theta\,(x_{n-1} - x_n) - F\cos\theta\, x_t & \text{if } x_n < \tan\theta \end{cases}$$

(12a)

Or

$$\begin{cases} \frac{dx_n}{dt} = \frac{1}{2}(F\cos\theta\, x_{n-1} + F\sin\theta) - F\cos\theta\,(x_n + x_t) & if\ x_n \geq \tan\theta \\ \frac{dx_n}{dt} = \frac{1}{2}F\cos\theta\,(x_{n-1} - x_n - 2x_t) & if\ x_n < \tan\theta \end{cases} \quad (12b)$$

This equation is subject to the boundary condition that $x_n$ becomes $l_{n-1}$ at $t = \tau$. At time zero, the length of the $n$-th terrace is $l_n$. We note that this equation introduces another variable $x_t$, whose governing equation is:

$$\begin{cases} \frac{dx_t}{dt} = F\cos\theta\,(x_n + x_t) & if\ x_n \geq \tan\theta \\ \frac{dx_t}{dt} = \frac{1}{2}F\cos\theta\,(x_n + 2x_t + \tan\theta) & if\ x_n < \tan\theta \end{cases} \quad (13)$$

This equation is subject to the boundary condition that $x_t$ becomes $l_n$ at $t = \tau$. At time zero, the length of the top terrace is a fixed value of zero – this may appear as the second or extra boundary condition to the first order differential equation, Eq. (13). As will become clear later – from Eq. (20) – this seemingly extra boundary condition is automatically satisfied under quasi-steady state and is therefore of no concern.

In principle, solving Eqs. (11-13) will give us $l_i$ for $4 \leq i \leq n$, in terms of deposition parameters ($\theta$ and $F$), as well as two parameters $\tau_0$ and $\tau$. We next establish two additional equations for $\tau_0$ and $\tau$. One equation is based on the critical nucleation condition, as given in Ref. [25]. Specifically, the average time of nucleating a new layer, or the period $\tau$ as used in this paper, for a given nucleation probability $P$ is:

$$\tau = \langle t \rangle = \frac{\int_0^{+\infty} t \frac{dP(t)}{dt} dt}{\int_0^{+\infty} \frac{dP(t)}{dt} dt} \quad (14)$$

The nucleation probability $P$ is [25]:

$$P(t) = 1 - e^{-\frac{4F^2 \cos^2\theta}{v_{2D}} \int_0^t x_t^3(\xi) d\xi} \quad (15)$$

The other equation is based on mass conservation. Specifically, the amount of materials deposited on an effective length of $L_s$, which is the separation of nanorods, all goes to the growth of terraces for the same period of time $\tau$. That is,

$$\tau F \cos\theta\, L_s = 2(l_1 + l_2 + \cdots + l_n) \quad (16)$$

Solving Eqs. (11-15) – as well using the Eqs. (3), (7) and (10) – we can obtain numerical values of all $l_i$'s as a function of deposition parameters ($\theta$, $F$ and $v_{2D}$ or temperature $T$). The sum on the right side of Eq. (16) also gives the nanorod diameter $L_m$.

In numerically solving these equations, for a set of deposition conditions – deposition rate, incidence angle, and substrate temperature or diffusion jump rate – we first determine $L_s$ [24], and use $L_s$ to make an initial estimate of $\tau$ as $(1-\epsilon)/F\cos\theta$, with $\epsilon$ being a small quantity 0.1 corresponding to $L_m/L_s = 0.9$. Further, we make an initiate estimate of $\tau_0$ as identical to $\tau$. Based on the initial estimates of $\tau_0$ and $\tau$, we can now solve for $l_1, l_2, \ldots, l_n$ with $n$ increasing until the following condition is satisified:

$$\begin{cases} 2(l_1 + l_2 + \cdots + l_{n-1}) \leq L_m \\ 2(l_1 + l_2 + \cdots + l_{n-1} + l_n) > L_m \end{cases} \quad (17)$$

With $n$ determined, we can then vary $\tau_0$ so that the following condition is satisfied:

$$2(l_1 + l_2 + \cdots + l_{n-1} + l_n) = L_m \tag{18}$$

Based on the converged $l_1$, $l_2$, …, and $l_n$ (with relative error no larger than $10^{-5}$), we can determine $\tau$ according to the nucleation condition, Eq. (14). This value of $\tau$ now replaces the initial estimate of $\tau$, and starts the next iteration of calculating $\tau_0$, then $l_1$, $l_2$, …, and $l_n$, then $\tau$ again, until the value of $\tau$ converges to a relative error of $10^{-5}$ or smaller.

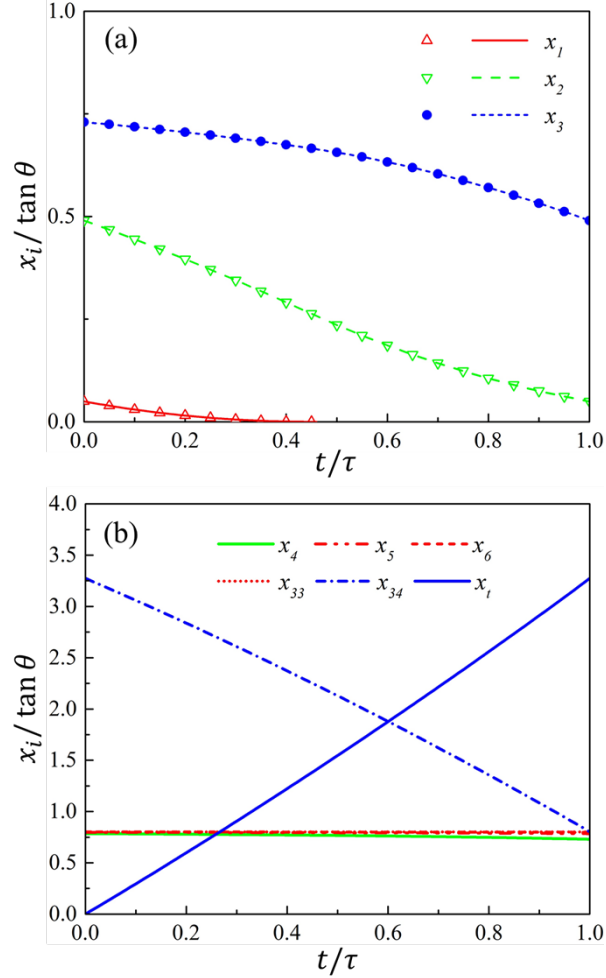

Fig. 3. Length of each terrace $x_i$, normalized by $\tan\theta$, as a function of time $t$, normalized by $\tau$; for (a) $x_1$, $x_2$ and $x_3$ with numerical solutions shown as lines and closed-form solutions as symbols; and for (b) $x_4$ to $x_{34}$. $x_7$ to $x_{32}$ are not shown since they overlap with $x_6$ and $x_{33}$. The length $x_t$ is also included for comparision.

To illustrate the variation of terrace lengths under quasi-steady state, we choose typical deposition conditions: 3 monolayer/s (ML/s) or about 1 nm/s as the deposition rate, 100 K (cooresponding to 300 K in three dimensions) as the substrate temperature, and 80° as the incidence angle. The nanorod separation $L_s$ is 365 in atomic unit according to Ref. [24]. Numerical solutions give $n = 34$, $\tau_0/\tau = 0.52$, and $L_m = 325$. As shown in Fig. 3(a), numerical solutions and closed-form theories of $x_1$, $x_2$ and $x_3$ in Eqs. (2), (6), and (9) are nearly identical; this indicates that the numerical solution procedure is reliable. As shown in Fig. 3(b), the lengths of terraces, from the 4th terrace to the 33rd terrace, are almost identical to each other and remain constant over time. Further, $x_i/\tan\theta$ is smaller than 1 for all the terraces except the top two, confirming that indeed deposition flux from left to right does not reach the terraces on the right side, except near the top.

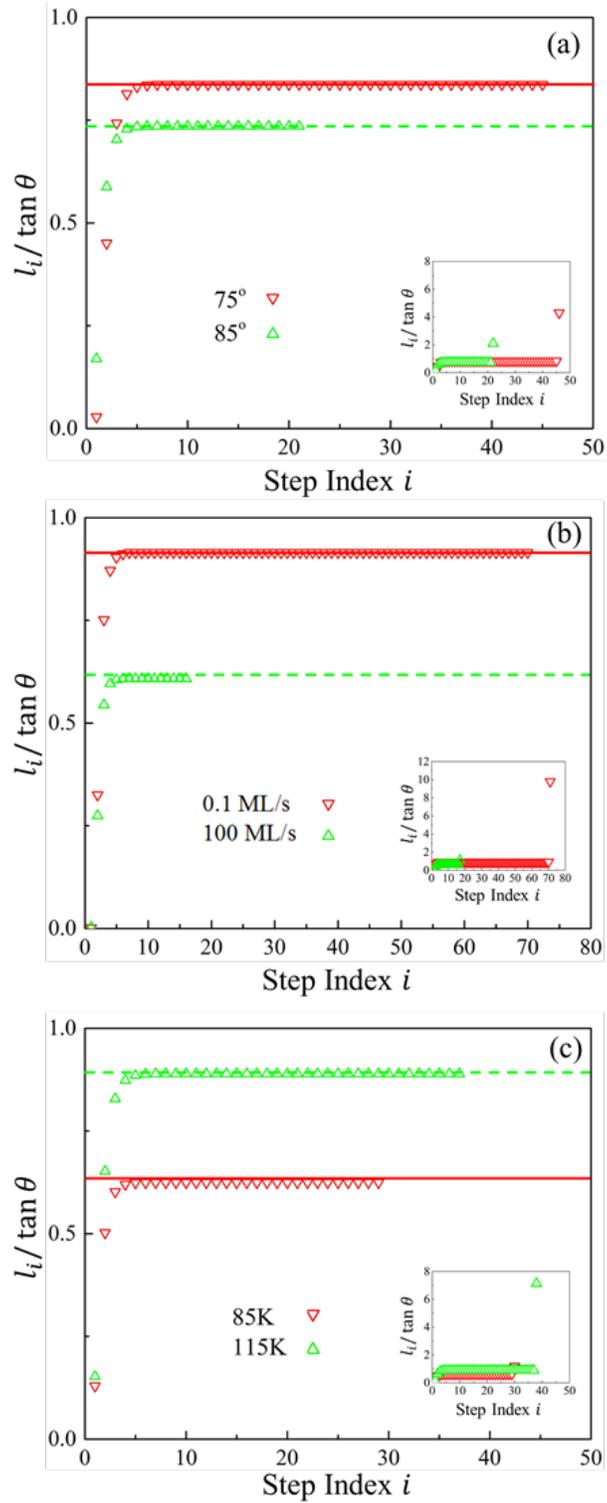

Fig. 4. Initial value of terrace length $l_i$, normalized by $\tan\theta$, as a function of step index $i$, (a) for two different incidence angles, (b) for two different deposition rates, and (c) for two different substrate temperatures. Symbols represent accurate numerical solutions, and lines represent closed-form theories.

Since $x_i$ remains constant over time and for different *i*'s ranging from 4 to 33 or (*n*-1), we next examine the initial value $l_i$ as a function of *i*, for various deposition conditions. Figure 4 shows the $l_i$ as a function of step index *i*, using the deposition conditions of Fig. 3 as reference. With the deposition rate of 3 ML/s and substrate temperature of 100 K, the constancy remains valid for two different incidence angles, and the constant value of terrace length increases as the incidence angle decreases but remains below $\tan\theta$; Fig. 4(a). With the incidence angle of 80° and substrate temperature of 100 K, the constancy remains valid for two different deposition rates, and the constant value of terrace length increases as the deposition rate decreases but remains below $\tan\theta$; Fig. 4(b). With the incidence angle of 80° and deposition rate of 3 ML/s, the constancy also remains valid for two different substrate terperaturse, and the constant value of terrace length increases as the temperature decreases but remains below $\tan\theta$; Fig. 4(c). Since terrace lengths are the same for different terraces, the overall morphology of the terraced surface is non-curved or a straight line in two dimensions. Further, the constancy of terrace lengths over time ensures that the straight line has a constant slope, which is the inverse of terrace length.

## 2.3 Closed-form Theories

One insight from the accurate numerical solutions in Section 2.2 is that the terrace length is constant over time and constant as terrace index *i* varies from 4 to (*n*-1). This insight makes it possible to arrive at closed-form theories of $l_i$'s and $L_m$.

Based on this insight, when *i* varies from 4 to (*n*-1), the *i*-th step advances due to two contributions: one from deposition on the *i*-th terrace $\frac{1}{2}F\cos\theta\, l_i$, and other from deposition on vertical step $\frac{1}{2}F\sin\theta \times 1$. Over the time period $\tau$, this step has advanced the distance of $l_i$. That is, $l_i = (F\cos\theta\, l_i + F\sin\theta)\frac{\tau}{2}$. Therefore:

$$l_i = \frac{F\sin\theta\tau}{2-F\cos\theta\tau} = \frac{\Theta}{2-\Theta}\tan\theta \quad \text{where } \Theta = F\cos\theta\,\tau \tag{19}$$

When compared with accurate numerical solutions, this closed-form theory of $l_i$ for $4 \leq i \leq (n-1)$ is accurate as shown in Fig. 4.

Based on the constant terrace length defined in Eq. (19), the initial length of the *n*-th terrace or the value of $l_n$ can be determined. Under the typical deposition conditions, $x_n$ is larger than $\tan\theta$ most of the time during growth; Fig. 3(b). Therefore, we use the condition $x_n \geq \tan\theta$ in Eqs. (12) and (13) to obtain:

$$x_t(t) = \frac{F^2\cos^2\theta\tan\theta\, t^2}{2(2-\Theta)} + F\cos\theta\, l_n t \tag{20}$$

$$l_n = x_t(\tau) = \frac{\tan\theta}{2(1-\Theta)(2-\Theta)}\Theta^2 \quad \text{where } \Theta = \tau F \cos\theta \tag{21}$$

Using Eq. (14), we can determine $\Theta$ as:

$$\Theta = \int_0^{+\infty} e^{-I(\xi)} d\xi \tag{22a}$$

$$I(\xi) = \left(\frac{\tan\theta}{l_{2D}}\right)^3 \frac{1}{(2-\Theta)^3}\left[\frac{\Theta^6\xi^4}{(1-\Theta)^3} + \frac{12\Theta^4\xi^5}{5(1-\Theta)^2} + \frac{2\Theta^2\xi^6}{1-\Theta} + \frac{4\xi^7}{7}\right] \tag{22b}$$

Here $l_{2D}$ is a characteristic length scale and it is defined as:

$$l_{2D} = 2\left(\frac{v_{2D}}{F\cos\theta}\right)^{\frac{1}{3}} \tag{23}$$

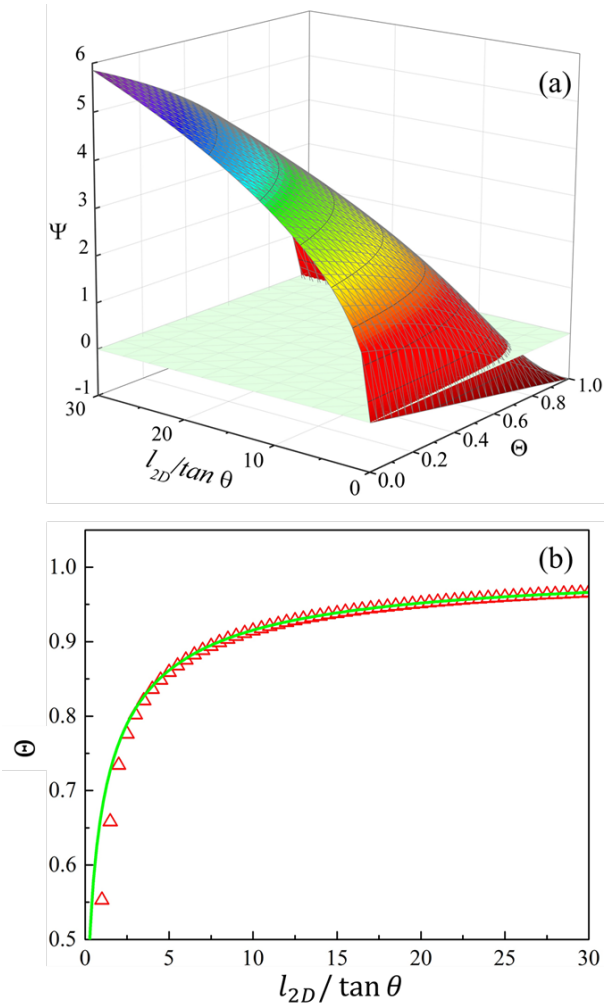

Fig. 5. (a) $\Psi$ as a function of $l_{2D}/\tan\theta$ and $\Theta$, and (b) $\Theta$ as a function of $l_{2D}/\tan\theta$ based on closed-form theory (solid line) and accurate numerical solution (triangle symbols) of $x_t$. The horoztonal plane at $\Psi = 0$ is included to highlight the $\Theta$ solution of Eq. (22).

In summary, the closed-form theories of $l_1$, $l_2$, and $l_3$ are given in Eqs. (3), (7), and (10); the closed-form theory of $l_i$, with $4 \leq i \leq (n-1)$, is given in Eq. (19); the closed-form theory of $l_n$ is given in Eq. (21); and the closed-form theory of $L_m$ is given as $\Theta L_s$. Here, $\Theta$ is defined in Eq. (22). Since all the closed-form theories rely on the solution of $\Theta$, we examine Eq. (22) in more depth. For this purpose, we introduce a new function $\Psi$:

$$\Psi\left(\frac{l_{2D}}{\tan\theta}, \Theta\right) = \int_0^{+\infty} e^{-I(\xi)} d\xi - \Theta \qquad (24)$$

As shown in Fig. 5(a), there is a unique $\Theta$ for a given $l_i/\tan\theta$, in order for $\Psi = 0$ to be satisfied. In addition, Fig. 5(b) shows that the $\Theta$ based on the closed-form theory of $x_t$ in Eq. (20) and that based on accurate numerical solution of $x_t$ are nearly identical.

## 2.4 KMC Verification of Closed-form Theories

In section 2.3, we have derived the following closed-form theories: $l_i = \frac{\Theta}{2-\Theta}\tan\theta$ for $4 \leq i \leq (n-1)$, and $L_m = \Theta L_s$; and compared them with accurate numerical solutions to prove the validity of approximations. The normalized terrace length $l_i/\tan\theta$ is $\frac{\Theta}{2-\Theta}$, which is a function of only $l_{2D}/\tan\theta$; and the normalized diameter $L_m/L_s$ is $\Theta$, which is again a function of only $l_{2D}/\tan\theta$. For verification, we will therefore compare the closed-form theories with KMC simulation results as a function of $l_{2D}/\tan\theta$. The KMC simulations are for a range of deposition conditions: depostition rate from 0.1 ML/s to 100 ML/s, substrate temperature from 85 K to 115 K, and incidence angle from 75° to 85°. For each set of deposition condition, we carry out 20 independent simulations and determine the average and standard deviations.

As the first verification, we examine the morphological evolution using one typical deposition condition - the deposition rate of 3 ML/s, the substrate temperature of 100 K, and the incidence angle of 80°. As shown in Fig. 6(a), the nanorod growth reaches a quasi-steady state once the nanorod is beyond 100 ML in height. The nanorod separation needs to be sufficiently large in order to accommodate a large number of monolayer surface steps, so as to reach the quasi-steady state. As the inset of Fig. 6(a) shows, if the separation is beyond 150, the normalized terrance length $l_i/\tan\theta$, averaged over independent simulations and terraces of constant length, converges to $0.80\pm0.02$, in comparison with 0.80 as the closed-form theory gives.

Going beyond the visual results of nanorod morphologies, we have carried out a series of KMC simulations for verification of the closed-form theories. In the first group of simulations, the deposition rate is 3 ML/s, the substrate temperature is 100 K, and the incidence angle varies between 75° and 85°; various $\theta$ in Figs. 6(b) and 6(c). In the second group, the incidence angle is 80°, and the substrate temperature is 100 K, and the deposition rate varies between 0.1 ML/s and 100 ML/s; various $F$ in Figs. 6(b) and 6(c). In the third group, the incidence angle is 80°, the deposition rate is 3 ML/s, and the substrate temperature varies between 85 K and 115 K; various $T$ in Figs. 6(b) and 6(c). For different deposition rates, substrate temperatures, and incidence angles, Figs. 6(b) and 6(c) show that the closed-form theories of $l_i/\tan\theta$ and $L_m/L_s$ are in good agreement with KMC simulation results.

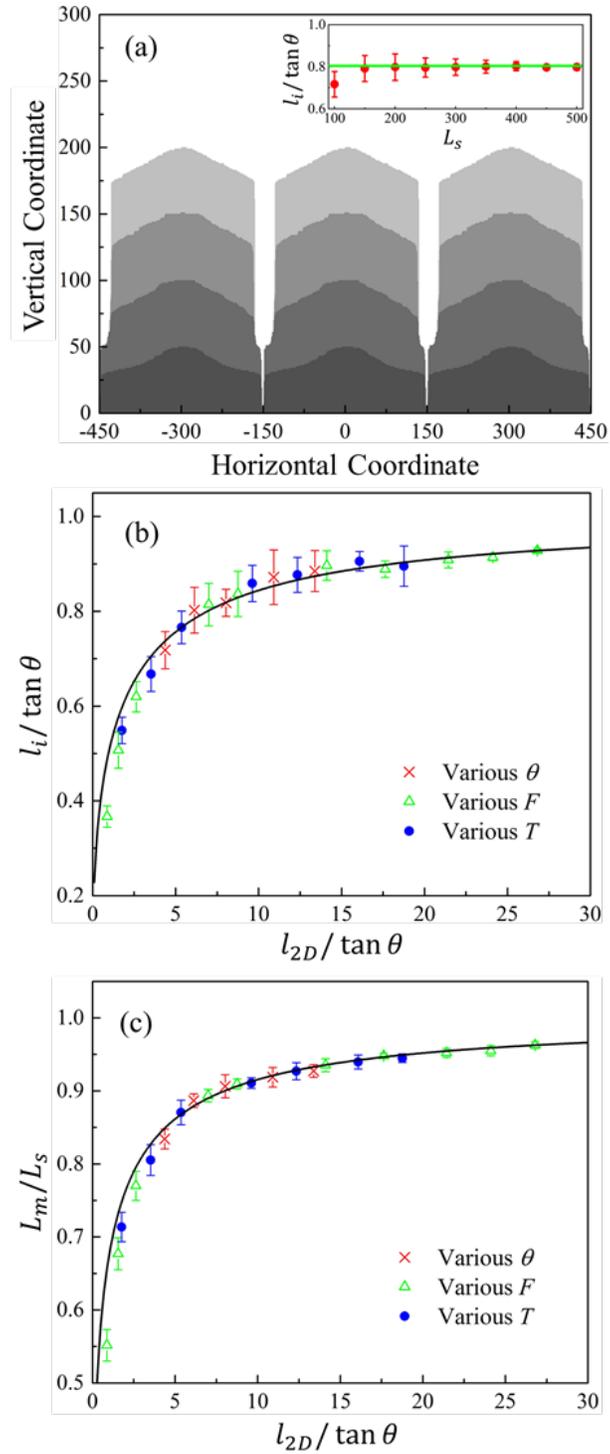

Fig. 6. (a) Morphologies of nanorods at four different height 50, 100, 150, and 200 ML, with the inset showing constant terrace lengths (or constant slope of terraced surfaces) as a function of nanorod separation; (b) $L_m/L_s$ and (c) $l_i/\tan\theta$ as a function of $l_{2D}/\tan\theta$, with closed-form theories shown as lines and KMC simulation results shown as symbols for various incidence angles $\theta$, deposition rates $F$, and substrate temperatures $T$.

## 3. Conclusions

We have developed closed-form theories of terrace lengths $l_i$ and nanorod diameter $L_m$ in growth mode I. In particular, the normalized terrace length $l_i/\tan\theta$ is $\frac{\Theta}{2-\Theta}$ for $4 \leq i \leq (n-1)$, and the normalized diameter $L_m/L_s$ is $\Theta$ during quasi-steady state growth; $l_1$, $l_2$, $l_3$, and $l_n$ are given in Eqs. (3), (7), (10) and (21). Here, $\Theta$ is defined in an integral form as a function of $\frac{l_{2D}}{\tan\theta}$. The complete set of closed-form theories is obtained based on the constancy of terrace lengths, and they have been proven accurate through comparison with numerical solutions without using the constancy condition. Further, the closed-form theories have been verified as valid by KMC simulations.

## Acknowledgements

The authors gratefully acknowledge the sponsorship of US Department of Energy Office of Basic Energy Science (DE-SC0014035).